# Structure in the disc of epsilon Aurigae: Spectroscopic observations of neutral Potassium during eclipse ingress

Robin Leadbeater, Three Hills Observatory, ( robin@threehillsobservatory.co.uk )

and

Robert E. Stencel, The Observatories, University of Denver ( rstencel@du.edu )

ABSTRACT
Variations in the equivalent width of the neutral potassium line at 7699A are reported, during ingress and into totality of the current eclipse of the enigmatic eclipsing binary epsilon Aurigae. The increase and plateaus of line strength are correlated with new system parameters and interferometric imaging constraints, plus ancillary data being reported contemporaneously. Together, these data reveal structural details of the transiting disc, never before measured. Measured ring and gap placements do not immediately fit any proposed model using simple tidal estimation. However, we predict egress times of interest and urge continued monitoring of this star during the balance of its eclipse, and encourage theoretical treatment of the disc substructure to be pursued.

1  Introduction

The bright eclipsing binary star, epsilon Aurigae, has long fascinated astronomers who could determine the nature of one companion star in the system, but not the other (Guinan and Dewarf, 2002; Stencel, 1985). The crux of the problem was that the apparent early F supergiant star in this single lined spectroscopic binary, should have a comparably massive secondary. However, the secondary is vastly underluminous for the presumed mass. Various models have been proposed, but the preferred model involves a secondary surrounded by a disc of material resulting from mass transfer, obscuring an otherwise normal companion star (Huang, 1965 & 1974).

In a study of spectroscopic data during the 1983 eclipse, Ferluga (1990) was able to demonstrate that the light curve and shell spectrum variations during eclipse were consistent with the eclipsing disc being defined as a system of rings seen nearly edge on. This included a central opening, radius 1.6 AU, a semi-transparent outer edge of the disc (radius 5.9 AU), and a transparent annular zone (radius 3.1 AU, width 0.8 AU) that splits the main opaque ring into two concentric bands. He found that the inclination of the disc plane with respect to the sky to be 82 degrees, and the full radial extent and height of the disc was 8.1AU by 0.9AU. These parameters can be compared with recent characterizations of the disc, based on (1) out of eclipse spectral energy distribution studies by Hoard, Howell and Stencel (2010) that describe the disc with radius 3.8 AU and thickness 0.95 AU, along with (2) new in-eclipse interferometric imaging (Kloppenborg et al. 2010) that confirmed the disc shape and motion during eclipse ingress. With additional interferometric imaging during totality, it is hoped that more disc parameters can be defined. Modern studies of generic circumstellar discs have made great progress in refining knowledge of their interior structure (cf. Benisty et al. 2010), but no other case of a transiting disc and its structure has been described before now.

In this report, we show that careful monitoring of the changing equivalent width of the K I line at 7699A during eclipse can be used to expand on details of disc substructure reported by Ferluga during the previous eclipse. This work allows this system to be placed in a larger context of disc dimensions and radial velocities, that begin to yield a consistent dynamical description of this dark companion.

2 Methods

2.1 Observations

Sixty four spectra of the neutral potassium line at 7699A were recorded at Three Hills Observatory between March 2009 and February 2010 using either a 200mm or 280mm aperture cassegrain telescope and a LHIRES III spectrograph modified to extend the range redwards. The plate scale was 7.7A/mm and resolution 0.3A (R~25000).

Epsilon Aurigae is circumpolar from the observatory located at 55 deg north allowing observations to be made throughout the year, though at high air mass (~6) around lower culmination. The target signal/noise ratio was > 100, though this was not achieved for some observations made at low elevation.

Data reduction was performed using IRIS and Visual Spec software:
http://www.astrosurf.com/buil/us/iris/iris.htm
http://www.astrosurf.com/vdesnoux/

The spectra were calibrated in wavelength using $O_2$ telluric lines as a reference, which are numerous in the region of the 7699A line (though fortunately far enough away from the line not to interfere) and strong from the near sea level altitude of the observatory. The telluric lines were removed by dividing by a spectrum of a bright star (either Altair or beta Aurigae) known to be line free in the 7699A region, measured at similar air mass. The spectra were normalized using a spline fit to the local continuum and corrected to the heliocentric reference frame. A list of spectra obtained is provided in Table 1 and a selection of spectra showing the development of the line profile with time is shown in Figure 1.

Measurements of the 7699A line outside eclipse by Hobbs (1974), Welty and Hobbs (2001) and Lambert and Sawyer(1986, hereafter L&S) show little variation in line strength, shape or radial velocity. Welty and Hobbs used high resolution spectroscopy to resolve several narrow components in the line, concluding that the component outside eclipse is interstellar in origin (Figure 2). The additional component which appears in the 7699A line during eclipse was therefore extracted by dividing each spectrum by the mean of the 8 spectra taken before the start of the eclipse (18 March-12 May 2009). The excess Equivalent Width (hereafter, eEW) due to this additional component was calculated (Table 1) and plotted against time (Figure 3). Note the conspicuous plateaus appearing during eclipse ingress.

The line was also divided into 0.3A wide zones and the eEW in each zone calculated. These are summarized in Table 2 and plotted against time in Figure 4. Figure 4 clearly shows the bulk of eEW develops in a zone centered near +23 km/sec, with contributions at smaller and larger speeds. We demonstrate in the next section that these variations can be understood in the context of Keplerian motion in the disc. This analysis also enables predictions for what might be expected during and after mid-eclipse.

The maximum radial velocity was estimated from the red edge of the line for each spectrum. (Table 1) and compared with the expected RV for a Keplerian disc orbiting a central star of mass 5.9M, as proposed by HHS (Figure 5). The agreement is good.

3  ANALYSIS

Interpretation of the discontinuous, step functions apparent in the K I equivalent width versus time data during ingress, is complicated by the fact that our line of sight view to the F star contains paths through the disc that may have contributions from interior structure. As Lambert and Sawyer (1986) noted, the K I lines are nearly but not quite optically thin during eclipse.

Figure 3 illustrates the step functions increases detected in the eEW of K I 7699A, which we summarize in Table 2 in terms of durations and slopes of the eEW changes. Adopting orbital motions during this portion of the eclipse, averaging 22 km/sec, we can translate the observation dates into relative displacements. Given the most dramatic rise in eEW begins close to the time of first contact, RJD 55,060 +/- 1 day (Santangelo, 2009 http://adsabs.harvard.edu/abs/2009ATel.2224....1S ), we adopt that as a reference time and compute the offsets in AU from that location. The sense of the observations is that eEW rises each time the relative motion of the disc and F star bring a layer of enhanced density into the line of sight. These are summarized in Table 3, where, for convenience, we denote these segments in order of appearance, as A, B, C, etc. Note that two of these segments, A and B, actually preceed first contact.

How are we to interpret the stepwise changes in excess equivalent width in the K I 7699A line? Perhaps the simplest explanation assumes that similar excitation conditions prevail across this portion of the disc, and thus changing density could give rise to the measured behavior. This interpretation suggests that structures akin to "rings" and "gaps" were detected in the disc plane. It would require a somewhat special optical depth, to enable inner nested rings to be seen transparently through outer ones, through the most opaque portions of the disc. Given the V band eclipse depth, backed by the reported opaqueness of the interferometric image obtained in H band (1.65 micron) continuum light (Kloppenborg et al. 2010), these enhancements in K I must arise from a more transparent 'atmosphere' above/below the disc and seen in absorption against the F star photosphere.

Surprisingly, two of these "rings" - which we will label A (RJD 54,976) and B (RJD 55,033) - appear well before V band eclipse began (RJD 55,060). Subsequent rings - C and D - showed much larger increase of excess equivalent width per day than the first two, and correlate well with the V band light curve reported by Jeff Hopkins in the Eclipse Campaign Newsletter series (http://www.hposoft.com/Campaign09.html ). These and others detected during ingress are presented in Table 3. Rings imply gaps, and gaps are associated with either tidal resonances (see below) and/or planet formation and zone clearing - an exciting prospect, if verification were possible.

In a Keplerian disc model, each rise in excess equivalent width can correspond to the leading edge of a new region of density enhancement, with extensions out of the disc plane. Ring-like and even spiral structure is suspected in an increasing number of situations including YSOs, Ae/Be and debris discs (see Hanuschik, 1994 and subsequent papers). For convenience, we will refer to these structures as rings, but caution the reader that these could be elliptical, spiral or even transient in character, varying from eclipse to eclipse.

It is useful to recognize that the presence of the F star can have at least three effects on the disc:
(1) heating the top and bottom of the disc, given that the F star diameter (1.5 AU) subtends more than the thickness of the disc (0.9 AU);
(2) heating of the facing portion of the disc, and
(3) tidal effects. Thus far, we are ignoring additional effects of the point-like (10 solar radii) unseen but presumed B5V star present in the center of the disc itself.

First, the F star radiation falling on the upper and lower surfaces of the disc could plausibly result in evaporation of volatile materials, including low excitation species like Na and K - seen in super-abundance during eclipse phases (cf. Lambert & Sawyer 1986). The Io sodium cloud around Jupiter, and similar enhancements in the tenuous atmosphere of Mercury due to solar wind bombardment of surfaces, provide an attractive analogy to this situation (see Schaefer and Fegley, 2009). However, if the K I enhancements are above and below the disc plane, either a radiative excitation instability is needed to produce the ring-like effect, and/or these are tracing an underlying density distribution. Both will require a theoretical treatment beyond the scope of this report.

Second, the heating of the face of the disc by the F star has been discussed previously (see Lissauer et al. 1996), where estimated temperatures rise to over 1000K. The strong asymmetry in ingress versus egress line strengths during eclipse, as discussed by Lambert and Sawyer (L&S, 1986) and other authors, combined with the disc rotation sense, suggest that strong optical depth effects are involved with the distance we can see into the disc, based on diagnostics like the K I excess equivalent width at different eclipse phases.

Third, what about tidal effects? We draw on the analogy of Kirkwood gaps in the asteroid belt caused by Jupiter, producing clearings among asteroidal orbits at ¼, 1/3, 2/5, 3/7, 1/2 and other integer fractions of the Jupiter orbital period. In epsilon Aurigae, the F star exterior to the disc plays the role of Jupiter. Adopting the mean F star mass, 2.75 solar, proposed by Hoard, Howell and Stencel (HHS, 2010), and the well known 27.1 year orbital period, one can predict the resonant locations for in the disc, ignoring other dynamical considerations, using Kepler's relation, $P^2 = a^3/M$, in year, AU and solar mass units. However, trial solutions require small integer fractions, such as ¼ P(orbit) to obtain a resonance at 5.03AU, near the A ring position, 1/8 to obtain 3.16AU near the CD gap, and then increasingly tiny ratios for more interior ones. To obtain gaps to fit the observed ones using the aforementioned Kirkwood gap ratios, the *effective* tidal mass needs to be minimal, e.g. 0.275 solar. At first look, this appears unsatisfactory.

Are the observed K I radial velocities consistent with the HHS mass for the central star in the disc? The Keplerian speeds for orbits at 2, 4 and 6 AU from the central star are 51, 36 and 30 km/sec, repsectively. The bulk of the gas velocities seen during ingress (Figure 4) occur at 23 km/sec, which relates to an orbit located 10AU from a 5.9 solar mass star. Curiously, this is approximately the dimension of the semi-major axis for the B star in the HHS model, suggesting the Roche lobe is filled with gas escaping from the disc. This suggests structure in K I 7699A may be detectable throughout much of the 27.1 year orbit cycle.

The red edge of the K I profiles trace a gas component velocity that appears consistent with gas tracing these Keplerian speeds through the disc as a function of eclipse phase (Figure 5). In Figure 4, the velocity substructure will be useful in diagnosing the locations of rings and gaps before AND after mid-eclipse. Among other facets, we note the emergence of blueward velocities after RJD 55,200, indicative of portions of the trailing portion of the disc starting to come into view (Figure 4).

Alternately, could a binary at the center of the disc cause these rings and gaps? HHS argue for a 5.9 solar mass B star at the center of the disc. If instead a pair of 3 solar mass stars were orbiting in the central clearing, 1AU radius, then the implied period would be 0.41 years. In that case, we predict 2/1, 3/1, 4/1, 5/1, 6/1, 7/1, 8/1... resonances at 1.6 (close to the EF gap), 2.1 (DE), 2.5 (?), 2.9 (CD), 3.3 (?), 3.6 (?), 3.9(BC)...AU. In this first-order estimate, we did not expect perfect correspondence, but rather wanted to illustrate the potential for independent mass constraints and dynamical studies possible with these data plus anticipated newer data with continued excess equivalent width monitoring of K I 7699A and related lines through the balance of eclipse this year and next. Subsequent study of these and other higher resolution data may reveal additional ring and gap components within the subset we have sampled, and these can further constrain possible dynamic features of the system.

The ring structure reported by Ferluga (1990) can be re-examined at this point. His contention was that a best fit model to the disc involved a semi-transparent (gas and dust) outer edge at 5.9AU from disc center, a ring gap 0.8AU wide centered at 3.1AU, plus a central hole starting at 1.6AU and inward. These dimensions were based on the high mass model, which can be scaled by 0.64, to match HHS and our application: 1.0AU (central opening), 1.9AU (our gap DE) and 3.8AU (disc radius per HHS), respectively. Ferluga and Mangiacapra (1991) calculated the radial extent of the disc to be 8.1 +/- 0.4 AU using shell spectral features, which scales to 5.0 — 5.5 AU -- consistent with the outermost rings we observed. As of this writing, the eclipse has reached the central clearing (late March 2010), another major dynamical structure, and monitoring this likely will provide additional information needed for a comprehensive disc model in the future.

For completeness, we include in Table 5 predictions for the egress phases of the current eclipse of epsilon Aurigae. If the rings and gaps are symmetric with respect to disc center, then evidence for these should be apparent at or near the predicted times. However, we note that predictions for egress times, based on prior eclipses, suggest that a more rapid emergence from eclipse is possible. Careful monitoring will reveal the rest of the story.

4 DISCUSSION

Spectroscopic monitoring of the neutral potassium line at 7699A has revealed step-like increases of equivalent width during ingress and early totality phases of epsilon Aurigae during 2009-2010. These can be interpreted as related to density enhancements inside, and surprisingly, outside of the denser disc causing the optical eclipses — the latter disc itself finally having been imaged interferometrically and its geometric parameters better established. An interesting future is forecast, both for dynamical modelling of the disc interior and atmosphere which extends apparently to the Roche limit of the secondary in the epsilon Aur system. We suggest the outer resonances mentioned above, should be observationally explored in K I and Na I lines between eclipses. Finally, whether these «rings» and «gaps» recur during egress will be watched with great interest — does their dynamical impetus prevail over the extra thermal input they receive when orbiting into the full UV glare of the F star primary? The orbital period for the disc material at 4AU from its presumed 5.9 solar mass central star, is 3.3 years, hence the material seen during egress (early 2011) already has been exposed to the F star since nearly the start of eclipse (2009).

We are grateful to Jeff Hopkins for his efforts in organizing the highly successful International Campaign for the observation of epsilon Aurigae during eclipse this cycle. R.E.S. is grateful for the bequest of William Herschel Womble to the University of Denver in support of astronomy, and for support under National Science Foundation grant # AST-1016678 to the University of Denver.

## TABLES:
### Table 1

| Date | JD - 2450000 | Signal/Noise | Total Excess Equivalent Width | Maximum Radial Velocity |
|---|---|---|---|---|
| | | | mÅ | km/s |
| 18-Mar-09 | 4909.38 | 105 | 4 | |
| 21-Mar-09 | 4912.42 | 135 | -10 | |
| 11-Apr-09 | 4933.37 | 90 | 5 | |
| 12-Apr-09 | 4934.36 | 150 | 12 | |
| 18-Apr-09 | 4940.35 | 140 | 2 | |
| 10-May-09 | 4962.38 | 85 | -3 | |
| 11-May-09 | 4963.38 | 90 | -12 | |
| 12-May-09 | 4964.37 | 105 | 3 | |
| 24-May-09 | 4976.52 | 60 | 25 | 24.9 |
| 29-May-09 | 4981.56 | 45 | 35 | 34.3 |
| 03-Jun-09 | 4986.50 | 60 | 20 | 28.8 |
| 05-Jun-09 | 4988.49 | 95 | 21 | 36.2 |
| 11-Jun-09 | 4994.50 | 110 | 21 | 27.3 |
| 25-Jun-09 | 5008.48 | 100 | 32 | 24.5 |
| 01-Jul-09 | 5014.46 | 50 | 40 | 37.8 |
| 19-Jul-09 | 5032.70 | 120 | 72 | 31.6 |
| 21-Jul-09 | 5034.57 | 65 | 89 | 32.0 |
| 30-Jul-09 | 5043.39 | 110 | 77 | 34.3 |
| 05-Aug-09 | 5049.55 | 145 | 78 | 29.2 |
| 07-Aug-09 | 5051.59 | 115 | 89 | 32.7 |
| 10-Aug-09 | 5055.06 | 100 | 80 | 32.3 |
| 20-Aug-09 | 5064.51 | 75 | 96 | 32.3 |
| 28-Aug-09 | 5072.46 | 65 | 128 | 32.7 |
| 08-Sep-09 | 5083.52 | 135 | 162 | 37.0 |
| 12-Sep-09 | 5087.49 | 130 | 169 | 32.7 |
| 15-Sep-09 | 5090.47 | 100 | 176 | 33.5 |
| 30-Sep-09 | 5105.48 | 85 | 244 | 39.0 |
| 03-Oct-09 | 5108.55 | 110 | 252 | 36.6 |
| 11-Oct-09 | 5116.45 | 95 | 301 | 39.4 |
| 16-Oct-09 | 5121.42 | 150 | 321 | 38.6 |
| 22-Oct-09 | 5127.45 | 130 | 319 | 38.2 |
| 28-Oct-09 | 5133.49 | 80 | 329 | 41.3 |

**Table 1 continued**

| Date | JD - 2450000 | Signal/ Noise | Total Excess Equivalent Width | Maximum Radial Velocity |
|---|---|---|---|---|
| | | | mÅ | km/s |
| 02-Nov-09 | 5138.44 | 90 | 327 | 41.7 |
| 05-Nov-09 | 5141.50 | 190 | 330 | 50.7 |
| 08-Nov-09 | 5144.45 | 130 | 320 | 40.9 |
| 14-Nov-09 | 5150.44 | 135 | 318 | 51.0 |
| 23-Nov-09 | 5159.34 | 100 | 315 | 40.5 |
| 26-Nov-09 | 5162.38 | 120 | 344 | 47.5 |
| 27-Nov-09 | 5163.25 | 130 | 341 | 41.3 |
| 30-Nov-09 | 5166.35 | 160 | 361 | 40.1 |
| 03-Dec-09 | 5169.46 | 95 | 348 | 42.1 |
| 07-Dec-09 | 5173.65 | 155 | 365 | 42.1 |
| 11-Dec-09 | 5177.27 | 135 | 380 | 47.5 |
| 18-Dec-09 | 5184.53 | 145 | 382 | 44.0 |
| 19-Dec-09 | 5185.56 | 230 | 390 | 49.9 |
| 26-Dec-09 | 5192.24 | 150 | 387 | 50.7 |
| 27-Dec-09 | 5193.57 | 135 | 382 | 49.9 |
| 28-Dec-09 | 5194.24 | 155 | 389 | 45.6 |
| 02-Jan-10 | 5199.26 | 165 | 395 | 49.1 |
| 05-Jan-10 | 5202.24 | 160 | 386 | 49.1 |
| 08-Jan-10 | 5205.24 | 175 | 397 | 48.7 |
| 17-Jan-10 | 5214.25 | 175 | 395 | 48.7 |
| 18-Jan-10 | 5215.24 | 200 | 393 | 49.5 |
| 27-Jan-10 | 5224.26 | 200 | 423 | 49.5 |
| 29-Jan-10 | 5226.29 | 190 | 415 | 52.2 |
| 31-Jan-10 | 5228.28 | 245 | 420 | 53.4 |
| 02-Feb-10 | 5230.28 | 195 | 420 | 52.6 |
| 06-Feb-10 | 5234.27 | 190 | 420 | 53.0 |
| 09-Feb-10 | 5237.47 | 175 | 427 | 54.2 |
| 17-Feb-10 | 5245.29 | 215 | 465 | 58.4 |
| 18-Feb-10 | 5246.40 | 140 | 469 | 54.2 |
| 20-Feb-10 | 5248.32 | 215 | 457 | 54.9 |
| 22-Feb-10 | 5250.34 | 235 | 476 | 54.2 |

**Table 2**

| Date | JD - 2450000 | range Å | Excess Equivalent Width by Zone | | | | | |
|---|---|---|---|---|---|---|---|---|
| | | | 7698.5 7698.8 | 7698.8 7699.1 | 7699.1 7699.4 | 7699.4 7699.7 | 7699.7 7700.0 | 7700.0 7700.3 |
| | | RV km/s | -12 | 0 | +12 | +23 | +35 | +47 |
| | | | mÅ (absorption is positive) | | | | | |
| 18-Mar-09 | 4909.38 | | 2 | 1 | 2 | -1 | -2 | 2 |
| 21-Mar-09 | 4912.42 | | -1 | -2 | -5 | -5 | 2 | 0 |
| 11-Apr-09 | 4933.37 | | -1 | -2 | 1 | 7 | 6 | -6 |
| 12-Apr-09 | 4934.36 | | 0 | -2 | 3 | 5 | 5 | 0 |
| 18-Apr-09 | 4940.35 | | -2 | 1 | 3 | 0 | 3 | -4 |
| 10-May-09 | 4962.38 | | 0 | 3 | 0 | -2 | -4 | 0 |
| 11-May-09 | 4963.38 | | -1 | 2 | -9 | -5 | -2 | 4 |
| 12-May-09 | 4964.37 | | 5 | -2 | 4 | 1 | -7 | 2 |
| 24-May-09 | 4976.52 | | 2 | 2 | 6 | 12 | -4 | 7 |
| 29-May-09 | 4981.56 | | 6 | 1 | 10 | 11 | 1 | 6 |
| 03-Jun-09 | 4986.50 | | 7 | -3 | 9 | 8 | 2 | -3 |
| 05-Jun-09 | 4988.49 | | -1 | -1 | 8 | 11 | 6 | -1 |
| 11-Jun-09 | 4994.50 | | 3 | 3 | 8 | 9 | -4 | 3 |
| 25-Jun-09 | 5008.48 | | 4 | 1 | 9 | 10 | -2 | 10 |
| 01-Jul-09 | 5014.46 | | -1 | 8 | 19 | 9 | 4 | 1 |
| 19-Jul-09 | 5032.70 | | 0 | -2 | 36 | 29 | 7 | 3 |
| 21-Jul-09 | 5034.57 | | 1 | -3 | 39 | 33 | 12 | 7 |
| 30-Jul-09 | 5043.39 | | 3 | 2 | 41 | 28 | 4 | -1 |
| 05-Aug-09 | 5049.55 | | 2 | 3 | 42 | 30 | 3 | -1 |
| 07-Aug-09 | 5051.59 | | -1 | 3 | 49 | 33 | 4 | 1 |
| 10-Aug-09 | 5055.06 | | 2 | 1 | 49 | 27 | 2 | -1 |
| 20-Aug-09 | 5064.51 | | 1 | 4 | 56 | 36 | 0 | 0 |
| 28-Aug-09 | 5072.46 | | 3 | 0 | 75 | 46 | 6 | -2 |
| 08-Sep-09 | 5083.52 | | 1 | 7 | 88 | 61 | 8 | -3 |
| 12-Sep-09 | 5087.49 | | 1 | 2 | 92 | 63 | 10 | 1 |
| 15-Sep-09 | 5090.47 | | -1 | 2 | 94 | 68 | 11 | 2 |
| 30-Sep-09 | 5105.48 | | 0 | 14 | 122 | 88 | 20 | 1 |
| 03-Oct-09 | 5108.55 | | 1 | 15 | 124 | 90 | 20 | 4 |
| 11-Oct-09 | 5116.45 | | 1 | 22 | 149 | 101 | 23 | 4 |
| 16-Oct-09 | 5121.42 | | -1 | 28 | 157 | 107 | 25 | 5 |
| 22-Oct-09 | 5127.45 | | 2 | 27 | 163 | 100 | 27 | 1 |
| 28-Oct-09 | 5133.49 | | 0 | 38 | 151 | 108 | 27 | 4 |

**Table 2 continued**

| Date | JD - 2450000 | range Å | Excess Equivalent Width by Zone | | | | | |
|---|---|---|---|---|---|---|---|---|
| | | | 7698.5 7698.8 | 7698.8 7699.1 | 7699.1 7699.4 | 7699.4 7699.7 | 7699.7 7700.0 | 7700.0 7700.3 |
| | | RV km/s | -12 | 0 | +12 | +23 | +35 | +47 |
| | | | mÅ (absorption is positive) | | | | | |
| 02-Nov-09 | 5138.44 | | 4 | 49 | 155 | 98 | 21 | 0 |
| 05-Nov-09 | 5141.50 | | 7 | 61 | 154 | 94 | 15 | 0 |
| 08-Nov-09 | 5144.45 | | 2 | 38 | 166 | 93 | 21 | 0 |
| 14-Nov-09 | 5150.44 | | 4 | 39 | 164 | 84 | 24 | 2 |
| 23-Nov-09 | 5159.34 | | 3 | 41 | 157 | 86 | 29 | 0 |
| 26-Nov-09 | 5162.38 | | 3 | 44 | 173 | 91 | 30 | 2 |
| 27-Nov-09 | 5163.25 | | 3 | 44 | 171 | 93 | 30 | -1 |
| 30-Nov-09 | 5166.35 | | 3 | 47 | 174 | 96 | 39 | 2 |
| 03-Dec-09 | 5169.46 | | 1 | 61 | 166 | 90 | 26 | 2 |
| 07-Dec-09 | 5173.65 | | 3 | 42 | 172 | 101 | 42 | 6 |
| 11-Dec-09 | 5177.27 | | 3 | 61 | 172 | 97 | 38 | 9 |
| 18-Dec-09 | 5184.53 | | 5 | 56 | 172 | 96 | 44 | 8 |
| 19-Dec-09 | 5185.56 | | 6 | 66 | 180 | 96 | 37 | 5 |
| 26-Dec-09 | 5192.24 | | 4 | 59 | 176 | 99 | 45 | 5 |
| 27-Dec-09 | 5193.57 | | 5 | 67 | 173 | 96 | 37 | 4 |
| 28-Dec-09 | 5194.24 | | 9 | 64 | 174 | 96 | 38 | 8 |
| 02-Jan-10 | 5199.26 | | 7 | 65 | 172 | 96 | 45 | 9 |
| 05-Jan-10 | 5202.24 | | 8 | 68 | 167 | 94 | 43 | 7 |
| 08-Jan-10 | 5205.24 | | 7 | 74 | 179 | 95 | 39 | 3 |
| 17-Jan-10 | 5214.25 | | 7 | 73 | 169 | 93 | 44 | 9 |
| 18-Jan-10 | 5215.24 | | 8 | 68 | 170 | 93 | 45 | 9 |
| 27-Jan-10 | 5224.26 | | 15 | 88 | 173 | 98 | 41 | 8 |
| 29-Jan-10 | 5226.29 | | 12 | 75 | 163 | 103 | 48 | 14 |
| 31-Jan-10 | 5228.28 | | 16 | 83 | 162 | 100 | 47 | 12 |
| 02-Feb-10 | 5230.28 | | 16 | 79 | 159 | 104 | 48 | 14 |
| 06-Feb-10 | 5234.27 | | 19 | 85 | 155 | 102 | 46 | 13 |
| 09-Feb-10 | 5237.47 | | 17 | 80 | 155 | 108 | 51 | 16 |
| 17-Feb-10 | 5245.29 | | 20 | 85 | 155 | 119 | 61 | 25 |
| 18-Feb-10 | 5246.40 | | 26 | 83 | 152 | 119 | 61 | 27 |
| 20-Feb-10 | 5248.32 | | 24 | 79 | 150 | 117 | 62 | 24 |
| 22-Feb-10 | 5250.34 | | 23 | 87 | 151 | 123 | 66 | 28 |

Table 3. Summary of implied structure based on excess equivalent width changes in the KI 7699A profile during eclipse ingress.

| RJD= J.D. - 2,400,000 | Feature | AU from center |
|---|---|---|
| 54963 | A outer | 5.02 |
| 54969.5 | A | 4.94 |
| 54976 | A inner | 4.86 |
|  | Gap AB | 4.86 - 4.45 |
| 55008 | B outer | 4.45 |
| 55020.5 | B | 4.29 |
| 55033 | B inner | 4.13 |
|  | Gap BC | 4.13 — 3.80 |
| 55059 | C outer, 1st contact | 3.80 |
| 55090 | C | 3.40 |
| 55121 | C inner | 3.01 |
|  | Gap CD | 3.01 — 2.53 |
| 55159 | D outer | 2.53 |
| 55168 | D | 2.41 |
| 55177 | D inner | 2.30 |
|  | Gap DE | 2.30 — 1.81 |
| 55215 | E outer | 1.81 |
| 55219.5 | E | 1.75 |
| 55224 | E inner | 1.70 |
|  | Gap EF | 1.70-1.57 |
| 55234 | F outer | 1.57 |
| 55242 | F | 1.47 |
| 55250 | F inner | 1.37 |
| 55276 | Central clearing? | 1.0 |
| 55413 | Predicted mid-eclipse | 0 (position of F star West limb) |

Table 4: Adopted system parameters

| Component | Value | Reference |
|---|---|---|
| disc radius: | 3.8 AU | HHS, 2010 |
| Secondary star radius | 150Ro | Stencel et al., 2008 |
| Secondary star mass: | 5.9 Mo | HHS, 2010 |
| Adopted length of ingress: | 140 days | (RJD 55,060-55,200) |

Ring/gap radius, $P^2 = a^3/M$, implies that $a(n) = (nP)^{2/3} M^{1/3}$

where n is an integer fraction representing a resonance with the orbital period, 27.1 years.

Table 5: Predicted times during egress, 2010-2011 (RJD = J.D. - 2,400,000)

| Features | RJD | Calendar date |
|---|---|---|
| Adopted mid-eclipse | 55413 | 4 Aug 2010 |
| Disc east inner rim | 55550 | 19 Dec 2010 |
| F ring crossing | 55576 — 55592 | 14 — 30 Jan 2011 |
| E ring crossing | 55602 — 55611 | 9 — 18 Feb 2011 |
| 3$^{rd}$ contact, predicted | 55640 | 19 Mar 2011 |
| D ring crossing | 55649 — 55667 | 28 Mar — 15 Apr 2011 |
| 4$^{th}$ contact, predicted | 55695 | 13 May 2011 |
| C ring crossing | 55705 — 55767 | 23 May — 24 Jul 2011 |
| B ring crossing | 55793 — 55818 | 19 Aug — 13 Sep 2011 |
| A ring crossing | 55850 — 55863 | 15 — 28 Oct 2011 |

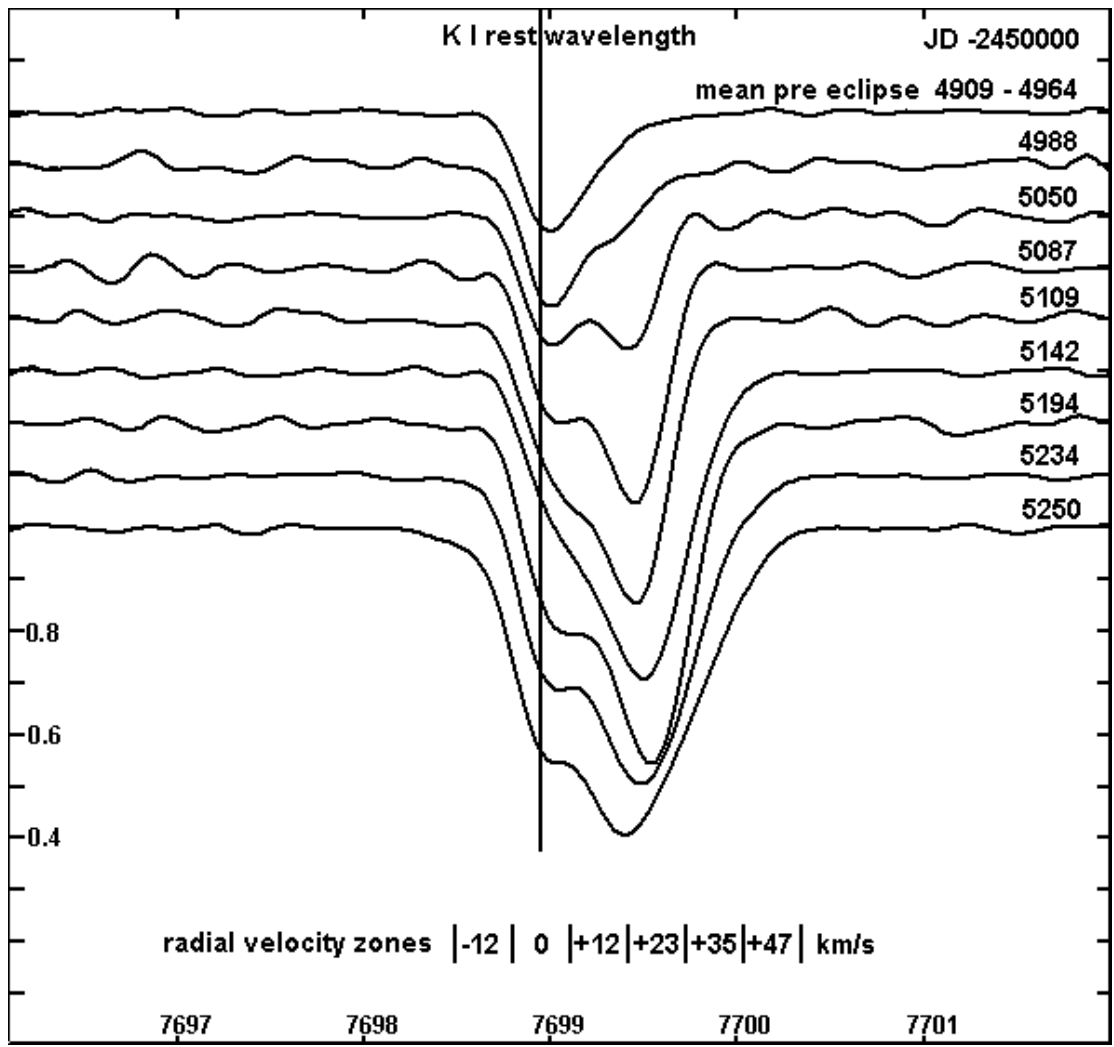

Figure 1. Time variation in the K I profiles during ingress and early totality.

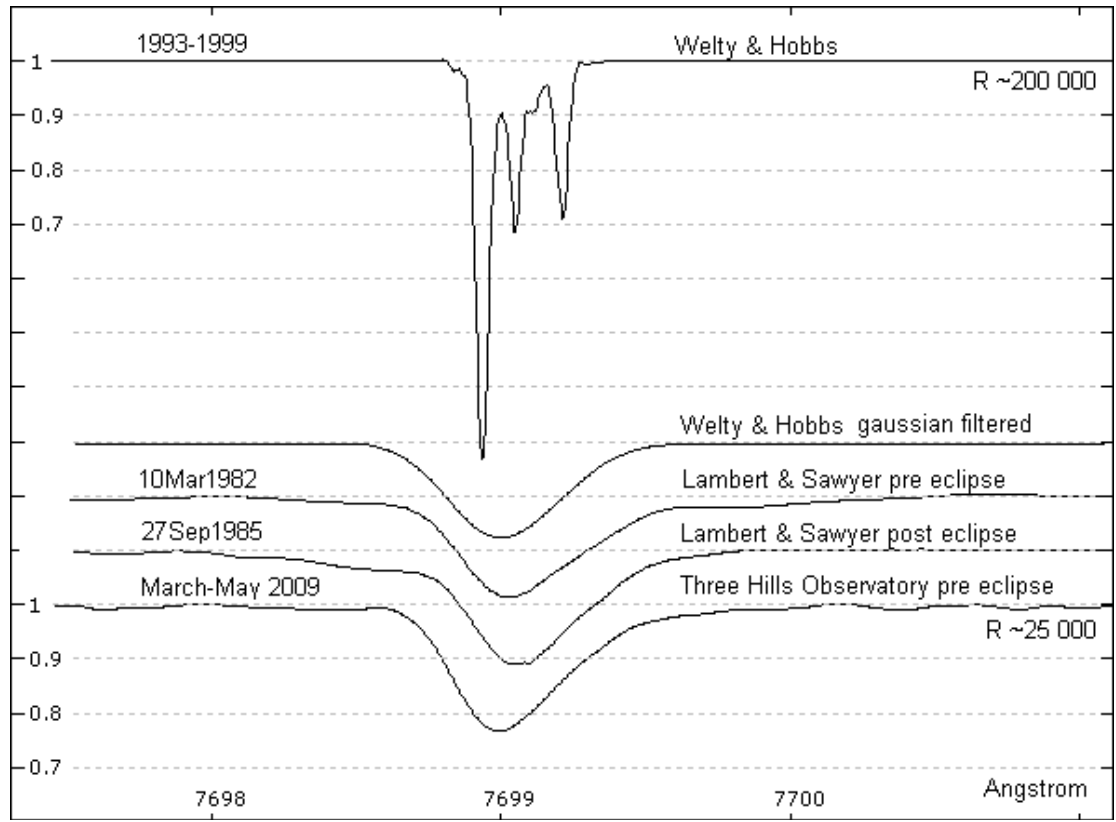

Figure 2. Pre-eclipse spectra compared with high resolution data from Welty and Hobbs (2001).

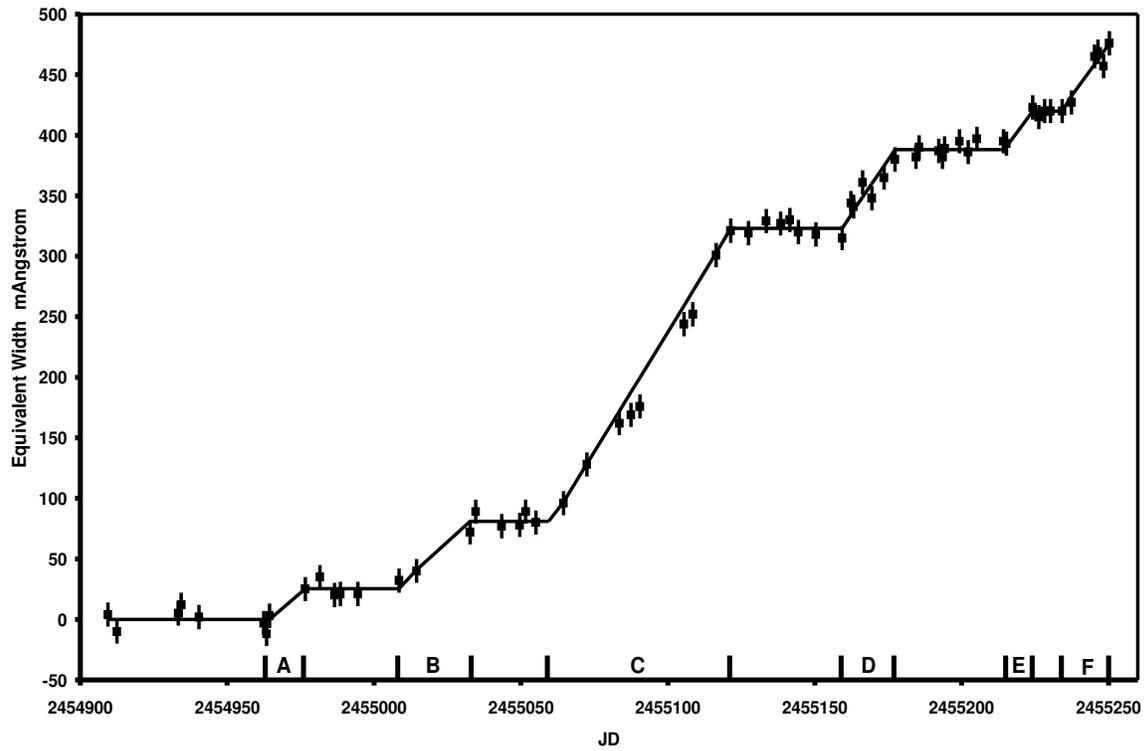

Figure 3. Time dependence of the excess equivalent width of the K I 7699A profile, minus the interstellar contribution, versus RJD, showing step functions during eclipse ingress and the start of totality.

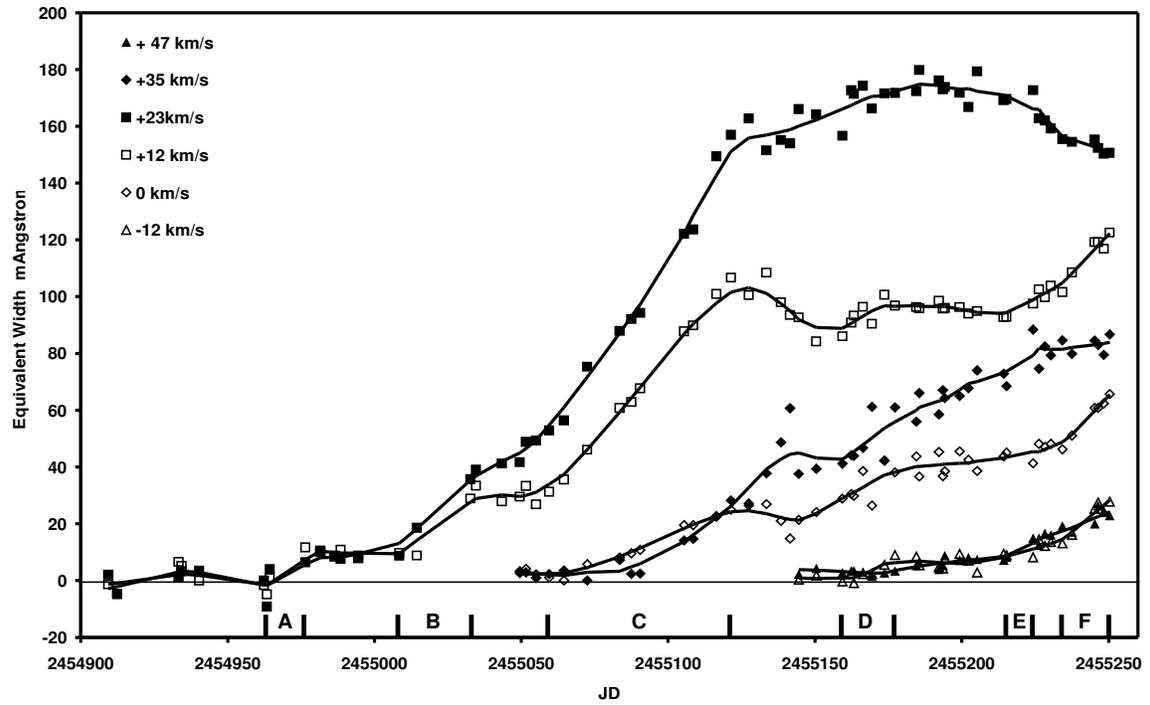

Figure 4. Variations in equivalent width of the KI profile as a function of radial velocity segment of the overall profile during ingress and part of totality.

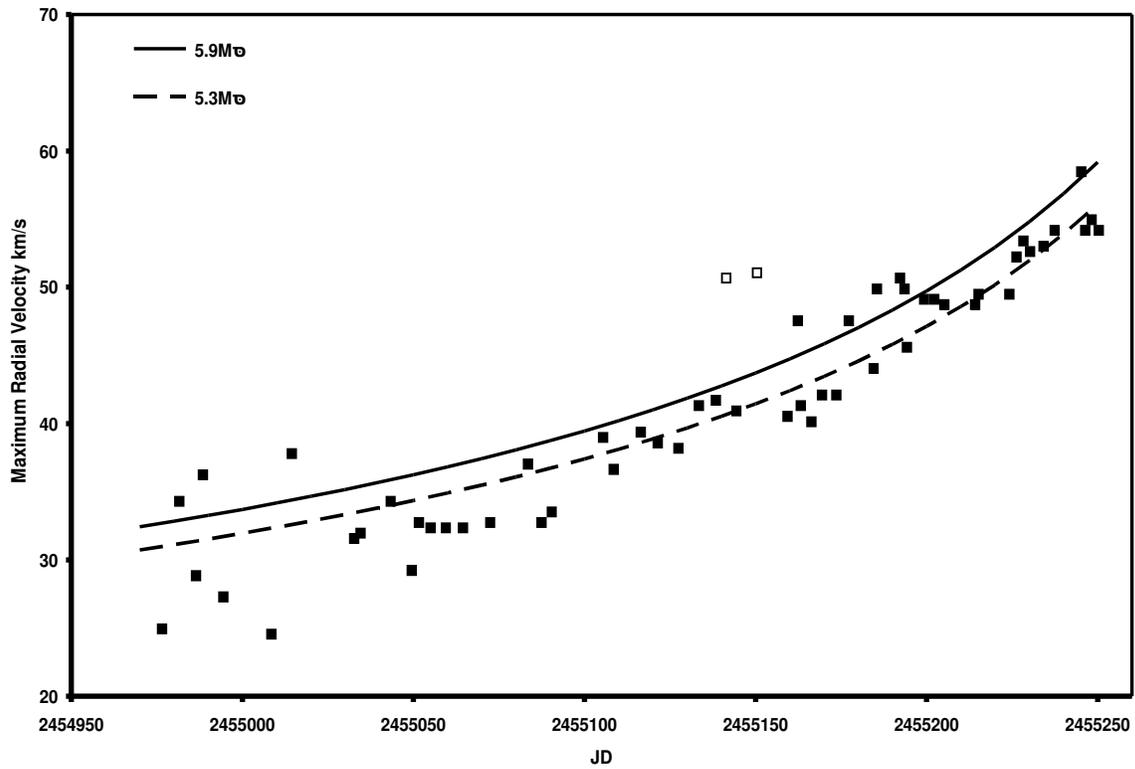

Figure 5. The maximum radial velocity, estimated from the red edge of the line for each spectrum. (Table 1), compared with the expected RV for a Keplerian disc orbiting a central star of 5.9 solar masses, as proposed by Hoard et al. (2010).

[end]